\documentstyle{elsart}

\input psfig
\input epsf

\begin{document}

\thispagestyle{empty}
~
\vspace{-27mm}

\hfill{\parbox[t]{3.5in}{\begin{flushright}
                                                RUB-TPII-19/95\\
                             \end{flushright} }}
\baselineskip=10pt

\vspace{0mm}

\begin{center}
{\LARGE\bf
Analytic structure of meson propagators\\
\vspace{0mm}
in the proper-time regularized\\
\vspace{5mm}
 Nambu--Jona-Lasinio model}

\vspace{6mm}
{\large Wojciech Broniowski$^a$, Georges Ripka$^b$, \\ \vspace{3mm}
 Emil N.~Nikolov$^c$, and Klaus Goeke$^c$}

\vspace{9mm}
{\sl $^a$ H. Niewodnicza\'nski Institute of Nuclear Physics,
PL-31342 Cracow, Poland}

\vspace{-2mm}
{\sl $^b$ Service de Physique Th\'eorique, Centre
d'Etudes de Saclay, F-91191 Gif-sur-Yvette Cedex, France}

\vspace{-2mm}
{\sl $^c$ Institut f\"ur Theoretische Physik II,
Ruhr-Universit\"at Bochum,\\D-44780 Bochum, Germany}

\end{center}

\vfill

\epsfxsize = 6 cm
\centerline{\epsfbox{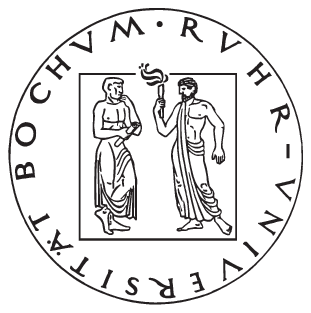}}
\vspace{0mm}

\baselineskip=7pt
\begin{center}
\Large Ruhr-Universit\"at Bochum \\
Institut f\"ur Theoretische Physik II\\
Teilchen- und Kernphysik
\end{center}

\newpage

\thispagestyle{empty}
~
\vspace{-25mm}

\hfill{\parbox[t]{3.5in}{\begin{flushright}
                                                RUB-TPII-19/95\\
                             \end{flushright} }}

\vspace{0mm}

\begin{center}
{\LARGE\bf
Analytic structure of meson propagators\\
\vspace{0mm}
in the proper-time regularized\\
\vspace{5mm}
 Nambu--Jona-Lasinio model}

\vspace{4mm}
{\large Wojciech Broniowski$^a$, Georges Ripka$^b$, \\ \vspace{3mm}
 Emil N.~Nikolov$^c$\footnote
{\parbox[t]{6in}{\baselineskip4mm On leave of absence from
Institute for Nuclear Research and Nuclear Energy, 1784 Sofia, Bulgaria;\\
DAAD fellow}},
 and Klaus Goeke$^c$}

\vspace{9mm}
{\sl $^a$ H. Niewodnicza\'nski Institute of Nuclear Physics,
PL-31342 Cracow, Poland}

\vspace{-1mm}
{\sl $^b$ Service de Physique Theorique, Centre
d'Etudes de Saclay, F-91191 Gif-sur-Yvette Cedex, France}

\vspace{-1mm}
{\sl $^c$ Inst. f\"ur Theoretische Physik II,
Ruhr-Universit\"at Bochum,
D-44780 Bochum, Germany}

\footnotetext[2]{E-mail:
\parbox[t]{5.5in}{\baselineskip4mm
wojtek@humboldt.ifj.edu.pl \\ ripka@amoco.saclay.cea.fr\\
emiln {\em or} goeke@hadron.tp2.ruhr-uni-bochum.de}}

\end{center}

\baselineskip=20pt
\parskip=0pt
\parindent=20pt

\vfill
\vfill
\noindent
We analyze the analytic structure of meson propagators in the
Nambu--Jona-Lasinio model with a proper-time regulator. We show that the
regulator produces unphysical complex singularities.
As a result the naive use of the
Wick rotation is no longer allowed.
Formulas involving integration over mesonic momenta,
such as meson-loop contributions or dispersion relations for meson
Green's functions, cannot be written in usual forms.

\vfill
\noindent
PACS: 12.39.-x, 12.39.Fe, 14.40.Aq

\vfill

\newpage

\setcounter{page}{1}

\section{Introduction}

\label{sec:intro}

The purpose of this paper is to discuss
analyticity properties of two-point functions
in the Nambu--Jona-Lasinio model. We point out
that the finite cut-off which regulates the quark loop can
lead to unphysical {\em complex} singularities in mesonic propagators.
This is the case of the proper-time regulator.
These complex singularities cause complications.
For instance, the analytic continuation from the Euclidean to the Minkowski
space has to be done with care, since the Wick rotation
picks up contributions from these poles (and also, as we will
show, from the infinite semi-circle).
Other regulators may be free of complex
poles ({\em e.g.} the Pauli-Villars case) but they have other problems.

The presence of complex poles is relevant
to calculations of hadronic properties in
effective low-energy models
(baryon calculations are reviewed {\em e.g.} in
Refs.~\cite{Wilets:rev,skyrmion:rev,BBC:rev,MCB:rev,Bo:rev,Tu:rev},
and meson ones in Refs.~\cite{Weise:rev,Vogl2:rev,Bijnens:rev}).
Properties of
baryons found at the soliton, or {\em mean-field} level, ({\em e.g.} baryon
mass, coupling constants, form factors, {\em etc.}) are leading-$N_c$
results. These quantities acquire meson-loop corrections,
which are $1/N_c$ suppressed. Since $N_c=3$,
some of these formally suppressed effects may be numerically large.
When Wick rotations are performed to calculate the meson-loop
contributions, the extra complex singularities contribute and may
not be neglected.

When discussing the regularization of effective theories,
such as the Nambu--Jona-Lasinio model,
it is useful to remember that regularization is very different from
renormalization, in which a cut-off is introduced and later taken to
infinity. The problems discussed in this paper disappear in the limit of
infinite cut-off. In the Nambu--Jona-Lasinio model, in order to fit data,
we are obliged to work with a finite
cut-off which rarely exceeds 1 GeV.
QCD does not provide us with a prescription for
regularizing the model.
Various regularizations have been used. The proper-time regularization of
the quark loop is most often used in soliton calculations because it
provides us with an action which is a local functional of the fields. This
simplifies soliton calculations as well as the definition of conserved
currents, which acquire familiar forms. Proper time regularization belongs
to the class of regularizations in which only the real part of the Euclidean
action is regularized. The reason for not regularizing the imaginary part is
two-fold. First, the proper-time regularization can only regularize a
positive definite operator, such as $DD^{\dagger }$, and it cannot be
applied to an operator such as $D/D^{\dagger }$, the log of which
contributes to the imaginary part of the Euclidean action\footnote{%
The imaginary part of the Euclidean action is odd in the $\epsilon^{\alpha
\beta \mu \nu}$ tensor, hence it describes anomalous processes. Conversely,
the real part, which does not contain the $\epsilon$ tensor, describes
non-anomalous processes. See {\em e.g.}
\cite{Zuk}.}. Second, 't Hooft's
anomaly matching condition, as well as phenomenology, instruct us that the
anomalous processes --- those which arise from the imaginary part of the
action --- should not be regularized.

However, other regularizations exist which yield the correct anomalies:
four-momentum regularizations, for example
\cite{Roberts:rev,Holdom,BallRipka}.
These can be expressed either in
terms of momentum-dependent constituent quark masses (or, equivalently,
non-local fields) or in terms of cut-off functions which multiply the quark
propagators. The non-locality of the resulting model yields extra terms
contributing to conserved currents. Four-momentum regularizations can and
should be applied to both the real and imaginary parts of the action. They
have the attractive feature that anomalous processes turn out to be
unregularized. In addition they
regularize the theory: both quark and mesons loops are regularized with a
single regulator. However, the non-locality induced by the four-momentum
cut-off regularizations make the soliton calculations an order of magnitude
longer. That is why the proper-time regularization is more often used and
why consider it in the present paper.

A further problem with meson loops is the need for introducing a new
cut-off. Indeed, the proper-time regularization will regularize the quark
loop but not the next-to-leading order meson loop corrections. The
regularization of the meson loop introduces an extra parameter in the model
to be fixed by some extra physical requirement.

\section{The regularized action}

\label{sec:NJL}

We work with the Lagrangian of the $SU(2) \times SU\left( 2\right)$-symmetric
Nambu--Jona-Lasinio model with scalar and pseudoscalar
interactions in the chiral
limit of vanishing current quark mass,
\begin{equation}
{\cal L}=\bar \psi i\partial ^\mu \gamma _\mu \psi +{\frac 1{2a^2}}\left(
(\bar \psi \psi )^2+(\bar \psi i\gamma _5{\vec \tau }\psi )^2\right) \;\;.
\label{eq:lagr}
\end{equation}
The partition function of the system is given by the path integral
\begin{equation}
{\cal Z}\equiv {\rm Tr}\;e^{-\beta H}={\frac 1{{\cal N}}}\int {\cal D}\phi
\;e^{-S(\phi )}\;\;,  \label{eq:Z}
\end{equation}
where $\phi $ is a four-component chiral field, $\phi =(\phi _0,\phi _1,\phi
_2,\phi _3)$, introduced to {\em partially bosonize} the model~\cite
{Eguchi:boson}, and ${\cal N}$ is a normalization factor. The Euclidean
action $S(\phi )$ consists of a quark part and a mesonic part,
\begin{equation}
S\left( \phi \right) =-{\rm Tr}\;\log D+{\frac{a^2}2}\int d^4x\phi ^2\;\;,
\label{eq:Seff}
\end{equation}
with the Dirac operator and the Dirac Hamiltonian defined as
\begin{equation}
D=\partial _\tau +h\;,\quad h=-i{\vec \alpha \cdot \vec \nabla }%
+\sum_a\Gamma _a\phi _a\;\;.  \label{eq:D}
\end{equation}
The quark-meson coupling matrices are $\Gamma _0=\gamma _0$ and $\Gamma
_i=i\gamma _0\gamma _5\tau _i$, $i=1,2,3$. The trace in Eq.~(\ref{eq:Seff})
runs over the quark field variables. The normalization factor in Eq.~(\ref
{eq:Z}) is equal to ${\cal N}=\int {\cal D}\phi e^{-\frac{a^2}2\phi ^2}$.

Most of the discussion of this paper concerns the so-called {\em proper-time}
regularization of the action~(\ref{eq:Seff}). Indeed it is one of the most
widely used regularizations, especially in connection with soliton
calculations \cite{Dyakonov86,Meissner90AP,Reinhardt88}. The proper-time
regulated action of the Nambu--Jona-Lasinio model is
\begin{equation}
S_\Lambda (\phi )={\frac 12} {\rm Tr} \int_{\frac 1{\Lambda ^2}}^\infty {%
\frac{ds}s}\,e^{-sD^{\dagger}D}+{\frac{a ^2}2}\int d^4x\phi ^2\ -{\frac 12}%
{\rm Tr}\;\log(D/D^{\dagger})\;.  \label{eq:sss}
\end{equation}
The first term is the real part of the action, which is regularized, whereas
the last term is the imaginary part of the action, which is finite and is
not regularized. In our $SU(2)$-case with no external sources the
contribution of the imaginary part vanishes and we drop it.

\section{Evaluation of the partition function}

\label{sec:loop}

We evaluate the partition function (\ref{eq:Z}) using the saddle-point
method. The stationary field configuration $\overline{\phi }$ is determined
by the equation:
\begin{equation}
\left. {\frac{\delta S(\phi )}{\delta \phi _a}}\right| _{\phi =\overline{%
\phi }}=0\;\;.  \label{eq:st-point}
\end{equation}
We consider the case of spontaneously broken chiral symmetry, where all the
fields $\bar \phi _a$ vanish except one which acquires the non-zero value: $%
\bar \phi _0=M$. We refer to $M$ as the constituent quark mass. Performing
the Gaussian integration over fluctuations of the field $\phi $ we get:
\begin{equation}
-\log {\cal Z}=S(\overline{\phi })+{\frac 12}{\rm Tr}\;\log (K^{-1}/a^2)\;\;,
\label{eq:Z1}
\end{equation}
where $K^{-1}$ is the inverse meson propagator, defined as
\begin{equation}
\left. \langle x,a\mid K^{-1}\mid y,b\rangle ={\frac{\delta ^2S(\phi )}{%
\delta \phi _a(x)\delta \phi _b(y)}}\right| _{\phi =\overline{\phi }}\;\;.
\label{eq:Kinv}
\end{equation}
The first term in Eq.~(\ref{eq:Z1}) is the {\em quark-loop} contribution and
it is proportional to $N_c$. It is the Hartree term. The second term in Eq. (%
\ref{eq:Z1}) is the {\em meson-loop} contribution, and it is proportional to
$N_c^0$. It contains both the Fock term and the RPA\ correlation energy. The
constant $a^2$ under the log results from the norm ${\cal {N}}$. Up to now
all of the calculations in the soliton sector have been limited to the
quark-loop level. In the vacuum sector a few attempts have been made to
include the meson-loop effects~\cite
{Cao92,Klevansky:loops,Zhuang94AP,Tegen95,loops:NJL}.

\section{Complex poles of the regularized meson propagators}

\label{sec:complex}

In this section we analyze the meson propagators $K$ in the vacuum sector
when the action is regularized by the proper-time method. We show that the
propagator has extra {\em complex} poles, in addition to the expected
singularities corresponding to the on-shell meson pole and the $q\overline{q}
$ production threshold. This feature has been overlooked in previous
investigations.

Since the chiral symmetry of the vacuum is spontaneously broken,
all the fields $\bar \phi _a$ vanish except one which acquires the
non-zero value: $\bar \phi _0=M$. We refer to $M$ as the constituent quark
mass. From~(\ref{eq:Kinv}) we deduce the following inverse pion and $\sigma $%
-meson propagators, which are functions of the Euclidean momentum $Q$:
\begin{equation}
K_{ab}^{-1}(Q^2)=4N_c\delta _{ab}(Q^2+\delta _{a0}\,4M^2)f(Q^2)\;\;.
\label{eq:explicit}
\end{equation}
The function $f(Q^2)$ is equal to:
\begin{equation}
f\left( Q^2\right) =\frac 1{16\pi ^2}\int_0^1du\int_{1/\Lambda ^2}^\infty
\frac{ds}se^{-s\left( M^2+Q^2u\left( 1-u\right) \right) }=\frac 1{16\pi
^2}\int_0^1duE_1\left( \frac{M^2+Q^2u\left( 1-u\right) }{\Lambda ^2}\right)
\label{eq:f}
\end{equation}
with $E_1(z)=\int_1^\infty dt/t\,e^{-zt}$. Equation (\ref{eq:explicit})
implies that the pion propagator ($a=1,2,3$) has a pole at $Q^2=0$, and that
the $\sigma $-meson propagator ($a=0$) has a pole at
\mbox{$Q^2=-4M^2$}.
These poles are interpreted as on-shell physical particles.
On the negative real axis in the $Q^2$ plane we find, in addition, a
quark-antiquark production threshold starting at $Q^2=-4M^2$, which is an
artifact of the lack of confinement. These are well known
features of the meson propagators in the Nambu--Jona-Lasinio model.

The meson
propagators have, however, additional singularities in the complex $Q^2$
plane. They are generated by the exponential regulator which appears in the
expression (\ref{eq:f}) of the function $f(Q^2)$. The regulator produces
extra {\em zeros} of the function $f\left( Q^2\right) $ in the complex $Q^2$
plane. This is displayed in Fig. \ref{fig:pionQ2}, which shows the analytic
structure of the pion propagator $K^{-1}$
in the complex $Q^2$ plane. We see that, in
addition to the physical pole and the $q\overline{q}$ cut, there are
infinitely many poles which are located close to the imaginary axis. The
figure is drawn for the special choice of parameters $\Lambda = M$
(which corresponds to some choice of the parameter $a$), but the
behavior is qualitatively the same for any other choice of parameters. The
location of first few poles is: $Q^2/\Lambda ^2=2.2\pm 18.0\,i$, $1.4\pm
41.7\,i$, $0.3\pm 66.6\,i$, {\em etc.} ($\Lambda = M$).
Asymptotically, for large $\mid
Q^2\mid $, the real parts of $Q^2$ at these poles tend to a constant, and
imaginary parts are separated by the value $8\pi \Lambda ^2$.

The result described above follows from a simple numerical calculation which
can be carried out {\em e.g.} with Mathematica. We just treat $Q^2$ as a
complex variable, and evaluate the integral over the $E_1$ function in
Eq.~(\ref{eq:f}) numerically. The function $E_1(z)$ has a cut
from 0 to $-\infty $, and is analytic elsewhere.
The integral in Eq.~(\ref{eq:f}) is well
behaved, and can be evaluated to any desired accuracy. The appearance of
complex zeros of the equation $f(Q^2)=0$ is not obvious at first glance, and
their location cannot be obtained analytically due to the transcendental
nature of the function.

A simple example can show why complex zeros appear in functions of this
type. Eq.~(\ref{eq:f}) involves averaging over exponential functions.
Consider a much simpler case of the function
\mbox{$g(z) = \int_0^1 d \alpha
\, e^{\alpha z} = z^{-1} (e^z - 1)$}. This function has zeros at locations $%
{\rm {Re}(z)=0}$, ${\rm {Im}(z)=2\pi k}$ (k = 0,1,...), which is very
similar to the case shown in Fig.~\ref{fig:pionQ2}. The function of Eq.~(%
\ref{eq:f}) is clearly more complicated, but the basic structure
remains. In Sec.~\ref{sec:twolev} we will encounter
another example of this behavior.

The structure such as shown in Fig.~\ref{fig:pionQ2} is
quite peculiar.
At first, it might seem that causality and unitarity are violated, since
singularities appear off the real negative axis in $Q^2$. However,
as demonstrated by Lee and Wick \cite{Lee-Wick}, and Cutkosky
{\em et al.} \cite{Cutkosky}, there exist models with complex singularities
which do satisfy causality and unitarity of the $S$-matrix.
This is achieved by
restricting the in and out scattering states to be composed of stable
modes only. Then unitarity
of the $S$-matrix can be maintained. For a two-point
function at low values of the $s$ variable,
the method of Cutkosky {\em et al.} is equivalent to calculate Feynman
diagrams of the perturbation theory along Wick-rotated
contours. At higher values of $s$ pinching of singularities may occur, and
the prescription is more involved.
Thus,
unitarity and (macro)causality
\cite{daniel:2} may be satisfied, despite the occurrence
of complex singularities.
An independent study would be required
to ascertain whether the prescription may be applied in our case.

Recently Langfeld and Rho \cite{Langfeld} analyzed a
confining Nambu--Jona-Lasinio-like model in which the pion propagator has,
in addition to the physical pole,
complex singularities located similarly
to our case.
In fact, complex singularities appear frequently in model
descriptions of confinement (for a review see \cite{Rob:conf}).

A related problem is the asymptotic behavior of the meson propagator for large
values of $\mid Q\mid ^2$ in the complex plane. In derivations of dispersion
relations, or in expressions for the one-meson-loop energy (see Sec.~\ref
{sec:energy}), one makes use of the Cauchy theorem. The unregularized meson
propagators behave asymptotically as $1/Q^2$, hence the integral over the
infinite semi-circle vanishes (after a sufficient number of subtractions),
and one obtains the usual dispersion relations. With the finite proper-time
cut-off, the asymptotic form of the meson propagator has a form involving an
exponential function, $\sim \exp (-\frac{Q^2}{4\Lambda ^2})h(Q^2)$, where $h$
is a slowly varying function. In applications of the Cauchy theorem, the
contribution from the infinite semi-circle does not vanish. This problem
will be discussed in more detail in Sec.~\ref{sec:twolev}.

We summarize the problems encountered in the Nambu--Jona-Lasinio model with
a proper-time regulator:

\begin{enumerate}
\item  The meson propagators
have (infinitely many) extra poles which lie
close to the imaginary axis in the complex $Q^2$
plane. After performing the 3-momentum integration, these poles unwind in
cuts in the energy variable.

\item  When performing a Wick rotation, the complex poles contribute. In
addition, we get contributions from the infinite semi-circle. As a result,
the usual dispersion relations for meson correlators and other Green's
functions acquire extra contributions.

\item  Note, however, that the pathological poles are quite far away from
the origin in the complex $Q^2$ plane. For the case of Fig.~\ref{fig:pionQ2}
(where $M=\Lambda $) the closest poles are about 18 units of $\Lambda ^2$
away, while the $q\overline{q}$ production threshold is 4 units of $\Lambda
^2$ away. Thus in some numerical calculations the presence of these poles
may not be immediately noticeable.

\item  In calculations of Green's functions at large values of
incoming momenta squared, a prescription like the one in Ref.~\cite{Cutkosky}
has to be used in order to satisfy basic requirements of field theory.

\end{enumerate}

\noindent
All above problems are due to the finiteness of the cut-off. In the limit $%
\Lambda \to \infty$ the poles at complex locations are moved to infinity,
and the usual analytic structure is recovered.

\section{Spectral representation of meson propagators in background soliton
fields}

\label{sec:spectral}

In this and the following sections, we limit the number of
quark states contributing to the spectral sum (\ref{eq:Ksol}) so as to
maintain the propagator $K$ finite even in the limit of infinite cut-off.
This allows us to obtain a better understanding of the extra poles which are
produced by the proper-time regulator.

We extend our results to the general
case with arbitrary background fields, {\em e.g.} for solitons.
We make use of the spectral representation, given by
the eigenstates of the Dirac Hamiltonian with the meson fields in a
saddle-point configuration:
\begin{equation}
h\mid j\rangle =\left( -i{\vec \alpha \cdot \vec \nabla }+\sum_a\Gamma
_a\bar \phi _a\right) \mid j\rangle =\epsilon _j\mid j\rangle \;,\quad
\langle \vec x\mid j\rangle =q_j(\vec x)\;\;.  \label{eq:spectrum}
\end{equation}
A straightforward calculation, outlined in App. \ref{app:der}, yields the
following expression for the inverse meson propagator
\begin{eqnarray}
&&\langle \vec x,a\mid K^{-1}(\nu )\mid \vec y,b\rangle =\delta (\vec x-\vec
y)\delta _{ab}a^2+\frac 1{4\sqrt{\pi }}\sum_{jk}\left( q_j^{\dagger }(\vec
x)\Gamma _aq_k(\vec x)\right) \left( q_k^{\dagger }(\vec y)\Gamma _bq_j(\vec
y)\right) \times  \nonumber  \label{eq:Ksol} \\
&&\int_{1/\Lambda ^2}^\infty \frac{ds}{\sqrt{s}}\left( -e^{-s\epsilon
_j^2}-e^{-s\epsilon _k^2}+s\left( (\epsilon _j+\epsilon _k)^2+\nu ^2\right)
\int_{-1}^1\frac{du}2e^{-\frac s4\left( 2(\epsilon _k^2+\epsilon _j^2)+\nu
^2+2(\epsilon _k^2-\epsilon _j^2)u-\nu ^2u^2\right) }\right) \;.  \nonumber
\\
&&
\end{eqnarray}
Note that the sum over the indices $j$ and $k$ runs over {\em all} the quark
states and that it includes a sum over color. In the infinite cut-off case
Eq.~(\ref{eq:Ksol}) reduces to the standard expression for the RPA
propagator \cite{Rowe,Ring,BRipka}:
\newpage
\begin{eqnarray}
&&\langle \vec x,a\mid K_{\Lambda \to \infty }^{-1}(\nu )\mid \vec
y,b\rangle =\delta (\vec x-\vec y)\delta _{ab}a^2-  \nonumber \\
&&\quad \sum_{ph}\left( \frac{\left( q_p^{\dagger }(\vec x)\Gamma _aq_h(\vec
x)\right) \left( q_h^{\dagger }(\vec y)\Gamma _bq_p(\vec y)\right) }{%
\epsilon _p-\epsilon _h+i\nu }+\frac{\left( q_h^{\dagger }(\vec x)\Gamma
_aq_p(\vec x)\right) \left( q_p^{\dagger }(\vec y)\Gamma _bq_h(\vec
y)\right) }{\epsilon _p-\epsilon _h-i\nu }\right) \;,  \label{eq:Ksolinf}
\end{eqnarray}
where the labels $p$ and $h$ denote empty particle ($\epsilon _p>0$) and
occupied hole ($\epsilon _h<0$) states.
(A finite quark chemical potential $%
\mu _q$ can be introduced so as to include possible valence states in the
set of occupied hole states.)
In expression (\ref{eq:Ksolinf}) the
sum is restricted to particle-hole excitations.
This follows ``automatically'' from Eq.~(\ref{eq:Ksol}). As
$\Lambda \to \infty$, the contributions to the spectral sum from pairs
of states with the same sign of the energy vanish identically.
This the expected Pauli-blocking effect.

\section{Energy of the soliton at the one-meson-loop level}

\label{sec:energy}

For the case where the background fields $\bar \phi $ are stationary, the
expression for the energy of the system is
\begin{equation}
E=-\frac 1\beta \log {\cal Z}=E_0+E_1-{\rm vac}\;\;,  \label{eq:E}
\end{equation}
where $E_0$ is the contribution obtained in the saddle-point approximation
containing the quark loop, and
$E_1$ is the contribution from the one-meson loop. The
vacuum energy (vac) is subtracted. The leading saddle-point contribution,
which is of order $N_c$ is:
\begin{equation}
E_0=\frac 1{4\sqrt{\pi }}\sum_j\int \frac{ds}{s^{3/2}}e^{-s\epsilon _j^2}+%
\frac{a^2}2\int d^3x\sum_a\bar \phi _a(\vec x)^2\;.  \label{eq:oneq}
\end{equation}
All soliton calculations up to now have been limited to this
leading-order contribution. In order to make contact with familiar
many-body theory, we define the quark-quark interaction:
\begin{equation}
V_{ij,lk}=-\int d^3x\,\sum_a\left( q_j^{\dagger }(\vec x)\Gamma _aq_i(\vec
x)\right) \frac 1{a^2}\left( q_k^{\dagger }(\vec x)\Gamma _aq_l(\vec
x)\right) \;.  \label{eq:Vijkl}
\end{equation}
Making use of Eq.~(\ref{eq:st-point}) we can cast the quark-loop
contribution in the form:
\begin{equation}
E_0=\frac 1{4\sqrt{\pi }}\sum_j\int_{\frac 1{\Lambda ^2}}^\infty \frac{ds}{%
s^{3/2}}e^{-s\epsilon _j^2}-\frac 12\sum_{jk}V_{jj,kk}\frac 1{4\pi
}\int_{\frac 1{\Lambda ^2}}^\infty \frac{ds}{s^{1/2}}\epsilon
_je^{-s\epsilon _j^2}\int_{\frac 1{\Lambda ^2}}^\infty \frac{ds^{\prime }}{%
s^{\prime }{}^{1/2}}\epsilon _ke^{-s^{\prime }\epsilon _k^2}\;\;.
\label{eq:oneq2}
\end{equation}
The second term is recognized as the Hartree energy shown diagrammatically
in Fig.~\ref{fig:qloop}. In the limit $\Lambda \to \infty ,$ the expression (%
\ref{eq:oneq2}) reduces to the familiar expression of the Hartree energy:
\begin{equation}
\lim_{\Lambda \to \infty }E_0=\sum_h\epsilon _h-\frac 12\sum_{hh^{\prime
}}V_{hh,h^{\prime }h^{\prime }}\;\;,  \label{eq:E0bis}
\end{equation}
The one-meson-loop contribution is
\begin{equation}
E_1=\frac 12\int_{-\infty }^\infty \frac{d\nu }{2\pi }{\rm Tr}\;\log \left(
K^{-1}(\nu )/a^2\right) =\frac 12\int_{-\infty }^\infty \frac{d\nu }{2\pi }%
\log \;{\rm Det}\left( K^{-1}(\nu )/a^2\right) \;\;,  \label{eq:E1}
\end{equation}
with $K^{-1}$ given by Eq. (\ref{eq:Ksol}). In the infinite cut-off limit we
may use the Cauchy theorem to evaluate the integral over $\nu $ in Eq. (\ref
{eq:E1}). As shown in App. \ref{app:log}, the integral of a $\log $ of a
(well behaved) function $f$ is proportional to the sum over all {\em zeros}
of $f$ in the upper complex plane, minus sum over {\em poles} of $f$ in the
upper complex plane. In our case $f$ equals to ${\rm Det}(K_{\Lambda \to
\infty }^{-1}/a^2)$. The zeros correspond to the eigen frequencies $\nu $
which satisfy the equation
\begin{equation}
{\rm Det}\,K_{\Lambda \to \infty }^{-1}(\nu )=0\;\;,  \label{eq:eigen}
\end{equation}
which is the usual RPA equation (see also App. \ref{app:RPA}). The poles of $%
{\rm Det}\,K_{\Lambda \to \infty }^{-1}$ are located at the particle-hole
excitation energies, as seen from Eq. (\ref{eq:Ksolinf}). As a result, we
obtain
\begin{equation}
\lim_{\Lambda \to \infty }E_1=\frac 12\int_{-\infty }^\infty \frac{d\nu }{%
2\pi }\log \;{\rm Det}\left( K_{\Lambda \to \infty }^{-1}(\nu )/a^2\right)
=\frac 12\sum_{\omega _i>0}\omega _i-\frac 12\sum_{ph}(\epsilon _p-\epsilon
_h)\;\;,  \label{eq:loopen}
\end{equation}
where
\begin{equation}
\omega _i={\cal I}m(\nu _i^0)  \label{eq:ome}
\end{equation}
In App.~\ref{app:RPA} we show that Eq.~(\ref{eq:loopen}) may be rewritten as
\begin{equation}
\lim_{\Lambda \to \infty }E_1=-\frac 12\sum_{i>0}\omega _i\mid Y_i\mid ^2-%
\frac{\langle P^2\rangle }{2M}+\frac 12\sum_{ph}V_{ph,hp}\;.
\label{eq:loopenfin}
\end{equation}
We recognize here the usual RPA expression, with a zero-mode contribution
\cite{Rowe,Ring,BRipka}.
The last term is the Fock
term, represented diagrammatically in Fig.~\ref{fig:mloop}(a). The ring
diagrams of Fig.~\ref{fig:mloop}(b-c) represent the RPA correlation energy.
Note that the meson-loop energy is not just the sum over the RPA frequencies
of the form $\frac 12\sum_i\omega _i$, as sometimes claimed \cite
{Tu:rev,Tu:mass}. The form of Eq.~(\ref{eq:loopenfin}) comes from the $q\bar
q$ substructure of the meson propagator. Indeed, in purely mesonic models
such as the Skyrme model, $E_1=\frac 12\sum_i\omega _i$ at the $N_c^0$ level.

As mentioned before, $E_1$ is suppressed 
by one power of $N_c$ compared to $E_0$. 
However, many-body physics provides 
examples of $1/N_c$-suppressed results which are
important. One effect is the so-called zero-mode correction, which accounts
for spurious contributions of the center-of-mass motion. The mean field
approximation used in soliton calculations does not separate the
center-of-mass coordinates from intrinsic coordinates. This notorious
problem of projecting out the center-of-mass motion of solitons has
attracted a lot of attention \cite
{Tu:mass,ZWM:mass,Dobado:mass,MK:mass,holz1:mass,holz2:mass,%
seattle:coh,GolliRosina} and numerical estimates give a negative
contribution of the order of 300 MeV. Some estimates \cite{Pob} use an
expression derived for systems with a finite number of particles in
the RPA approximation, as in Eq.~(\ref{eq:loopenfin})
\begin{equation}
\Delta M=-\frac{\langle P^2\rangle }{2M}\;\;  \label{eq:DM}
\end{equation}
where $P$ is the momentum operator, and $M$ the soliton mass. In general,
such a zero-mode subtraction appears for every continuous symmetry of the
Lagrangian which is broken by the solitonic solution. In addition to the
translational invariance, hedgehog solitons appearing in chiral models also
break the rotational and isospin symmetry, leaving only the sum of spin and
isospin preserved. In analogy to Eq. (\ref{eq:DM}) this leads to another
subtraction \mbox{$\Delta M = - \langle J^2 \rangle /(2 \Theta)$}, where $J$
is the angular momentum operator, and $\Theta $ is the moment of inertia
\cite{BCVector}. For typical model parameters, the rough estimates of the
zero-mode corrections are large, up to 30\% of the soliton mass, as expected.

In the derivation of Eq.~(\ref{eq:loopenfin}) the Wick rotation has been used.
As will be shown in next sections, extra contributions arise when the 
the proper-time regulator is present, and expressions as Eq.~(\ref{eq:loopenfin})
do not hold.

\section{Finite proper-time cut-off}

\label{sec:finite}
 We decompose the integrals over $s$ as follows:
\begin{equation}
\int_{1/\Lambda ^2}^\infty ds=\int_0^\infty ds-\int_0^{1/\Lambda ^2}ds\;\;,
\label{eq:sdec}
\end{equation}
and use
\begin{equation}
\int_0^{1/\Lambda ^2}\frac{ds}{\sqrt{s}}e^{-sr}={\frac{{{\sqrt{\pi }}\,{\rm %
erf}\left( {\frac{\sqrt{r}}\Lambda }\right) }}{{{\sqrt{r}}}}}\;,\quad
\int_0^{1/\Lambda ^2}ds\sqrt{s}e^{-sr}={\frac{{{\gamma }\left( {\frac 32},{%
\frac r{\Lambda ^2}}\right) }}{{{r^{{\frac 32}}}}}}=\frac{\sqrt{\pi }\gamma
^{*}\left( {\frac 32},{\frac r{\Lambda ^2}}\right) }{2\Lambda ^3}\;\;,
\label{eq:int2L}
\end{equation}
where the error function ${\rm erf}(z)$ and the incomplete Euler $\gamma $
functions $\gamma (n,z)$ and $\gamma ^{*}(n,z)$ are defined {\em e.g.}\ in
Ref.~\cite{AbrStegun}. We decompose the total contribution to $K^{-1}$
into the infinite cut-off part, and a remainder, which describes the
finite-cut-off effects:
\begin{equation}
K^{-1}(\nu )=K_{\Lambda \to \infty }^{-1}(\nu )+\Delta K^{-1}(\nu )\;\;,
\label{eq:diff}
\end{equation}
where
\begin{eqnarray}
\ &&\langle \vec x,a\mid \Delta K^{-1}(\nu )\mid \vec y,b\rangle =\frac
14\sum_{jk}\left( q_j^{\dagger }(\vec x)\Gamma _aq_k(\vec x)\right) \left(
q_k^{\dagger }(\vec y)\Gamma _bq_j(\vec y)\right) \times  \nonumber \\
&&\ \left[ -\left( \frac{{\rm erf}\left( \frac{\mid \epsilon _j\mid }\Lambda
\right) }{\mid \epsilon _j\mid }+\frac{{\rm erf}\left( \frac{\mid \epsilon
_k\mid }\Lambda \right) }{\mid \epsilon _k\mid }\right) \right. +  \nonumber
\\
&&\ \left. \frac{(\epsilon _j+\epsilon _k)^2+\nu ^2}{4\Lambda ^3}%
\int_{-1}^1du\,\gamma ^{*}\left( {\frac 32},\frac{\epsilon _k^2+\epsilon
_j^2+\nu ^2/2+(\epsilon _k^2-\epsilon _j^2)u-\nu ^2u^2/2}{2\Lambda ^2}%
\right) \right] \;\;.  \label{eq:pistar}
\end{eqnarray}
The function $\gamma ^{*}(a,z)$ is an entire function of the variable $z$ (%
{\em i.e.} it is single-valued and has no singularities in the finite
complex plane -- singularities occur only at infinity). Therefore $\langle
\vec x,a\mid \Delta K^{-1}(\nu )\mid \vec y,b\rangle $ is also an entire
function in the variable $\nu $. This is because $\nu $ enters in the
argument of $\gamma ^{*}$ in a non-singular combination with the integration
variable $u$. In other words, $\Delta K^{-1}(\nu )$ has no singularities in
the finite complex plane. As a result, the singularities of the function $%
K^{-1}(\nu )$ are generated by its $\Lambda \to \infty $ part in Eq.~(%
\ref{eq:diff}). This means they occur at particle-hole excitations,
$\nu = \pm i (\epsilon_p - \epsilon_h)$.

The zeros of ${\rm Det}K^{-1}(\nu )$ will of course be at different
locations than zeros of the $\Lambda \to \infty $ part. As we change $%
\Lambda $ from infinity to a finite value, the zeros of ${\rm Det}K^{-1}$
move away from those of ${\rm Det}K_{\Lambda \to \infty }^{-1}$.
In addition, new
unphysical zeros emerge at complex values of $\nu $. The situation is
analogous to the case of the vacuum meson propagators, discussed in Sec.~\ref
{sec:complex}.

\section{Two-level model}

\label{sec:twolev}

To illustrate how the new zeros of ${\rm Det}K$ are generated, we consider a
two-level model, in which just one particle state $p$ and one hole
state $h$ contribute to the spectral decomposition (\ref{eq:Ksol}) of the
propagator $K$. In this case only one mesonic fluctuation occurs. Fig.~\ref
{fig:cauchy} shows the analytic structure of the corresponding inverse meson
propagator $K^{-1}\left( \nu \right) $ in the complex plane of the energy
variable $\nu $. We note the poles (filled dots) located at  $\pm i(\epsilon
_p-\epsilon _h)$. The physical zeros (empty dots) on the imaginary axis
correspond to the RPA vibration energies. We also see infinitely many
unphysical zeros (empty dots) close to the axes at $45^o$ and $135^o$. If it
were drawn in the $\nu ^2$ complex plane, Fig.~\ref{fig:cauchy} would
look very similar to Fig.~\ref{fig:pionQ2}. The displayed structure
results from a numerical calculation, where we evaluate expression
(\ref{eq:Ksol}) for the two-level case.

We can carry out an even simpler analysis.
Notice that for large values of $%
\mid \nu \mid $, the function $K^{-1}$ in the two-level model can be
approximated by:
\begin{equation}
K^{-1}(\nu^2 )/a^2 \simeq 1 + b(\nu ^2)e^{\frac{-\nu ^2}{4\Lambda ^2}} \;\;,
\label{eq:appr}
\end{equation}
where $b(\nu^2)$ is a slowly varying function. Assuming $b(\nu^2)$ to be a real
positive constant, and breaking up $\nu^2$ into real and imaginary parts, $%
\nu ^2=R+iI$, the zeros of the function $K^{-1}$ occur at the
locations
\begin{equation}
I=4\Lambda ^2(2k+1)\pi \;\;,\;\;\;R=-4\Lambda ^2\log b\;\;,  \label{eq:imre2}
\end{equation}
where $k$ is an integer. The difference between the imaginary parts of
successive zeros is $8\pi \Lambda ^2$. If we integrate by parts, as in Eq.~(%
\ref{eq:logbp}), then the integral to be considered is
\begin{equation}
J=\int_{-\infty }^\infty d\nu \log (K^{-1}(\nu )/a^2)=-\int_{-\infty
}^\infty d\nu \,\nu \frac{d K^{-1}(\nu )/d\nu }{K^{-1}(\nu )}\equiv
\int_{-\infty }^\infty d\nu j(\nu )\;\;.  \label{eq:logbp2}
\end{equation}
Denoting $\nu =\mid \nu \mid e^{i\phi }$, we find that for large $\mid \nu
\mid $
\begin{eqnarray}
j(\nu )\sim 0\quad  &{\rm for}&\quad \cos (2\phi )>0\;\;,  \nonumber \\
j(\nu )\sim \frac{\mid \nu \mid ^2}{2\Lambda ^2}e^{2i\phi }\quad  &{\rm for}%
&\quad \cos (2\phi )<0\;\;.  \label{eq:nuc2}
\end{eqnarray}
This allows us to check if we can close the contour in order to evaluate $J$
via the Cauchy theorem. As a consequence of Eq.~(\ref{eq:nuc2}), the
integral of $j(\nu )$ along the contours $C_1$ and $C_3$ in Fig.~\ref
{fig:cauchy} vanishes (this is the case $\cos (2\phi )>0$), but the integral
along the contour $C_2$ (case $\cos (2\phi )<0$) {\em does not vanish}, and
is equal to $\sqrt{2}\,\mid \nu \mid ^3/(6\Lambda ^2)$. %
Hence the integral of $j$ over the contour $C_2$ goes as $\mid \nu \mid ^3$.

The result is that $J$ in Eq.~(\ref{eq:logbp2}) is an infinite sum of the
pole contributions, plus an infinite contribution from the contour
$C_2$. Since we know that $J$ is finite,
cancelations between the infinities from the pole and the
$C_2$ parts occur, and we are left with a finite
(but non-zero) result.
Certainly, the same cancelation of infinities takes place in the general
case of Eq.~(\ref{eq:E1}), when we evaluate it via the Cauchy theorem.

\section{Other regulators}

\label{sec:otherreg}

The analytic structure displayed in Figs.~\ref{fig:pionQ2} and~\ref
{fig:cauchy} is specific to the proper-time regulator. Other popular
regulators, such as the sharp Euclidean four-momentum cut-off or the
Pauli-Villars regulator do not lead to complex singularities.
The Pauli-Villars regularization consist of
introducing additional families of quarks (which have Bose
statistics). In the semi-bosonized form the action has the form
\begin{equation}  \label{eq:pv}
{\cal L} = -{\rm Tr}\;\log (-i\gamma \cdot \partial + \gamma_0 \Gamma_a
\phi_a ) + \sum_{p=1}^{N_p} c_p {\rm Tr}\; \log (- i \gamma \cdot \partial +
g_p \gamma_0 \Gamma_a \phi_a ) + \frac{a^2}{2} \int d^4x \phi^2 \;,
\end{equation}
where the the label $p$ labels the Pauli-Villars quark families, $N_p$ is
the number of these families, and the constants $c_p$ and $g_p$ are chosen
appropriately in order to cancel infinities \cite{bo:PV}. It is clear from
Eq.~(\ref{eq:pv}) that the corresponding meson propagators will have
singularities along the real negative $Q^2$ axis only. Apart from poles,
there is the cut going from the $q\overline{q}$ production threshold at $%
Q^2=-4 M^2$ to $Q^2=-\infty$, and additional cuts associated with the
production of the Pauli-Villars quarks, going from $Q^2=-4 g_p^2 M^2$ to $%
Q^2=-\infty$. Therefore the Nambu--Jona-Lasinio model with the Pauli-Villars
regulator obeys causality. Moreover, the usual analyticity structure allows
for the straightforward use of the Cauchy theorem, rotation of contours,
{\em etc.} In particular, this leads to the formal expression for the RPA
correlation energy, such as in Eq.~(\ref{eq:loopenfin}).
The problem with the Pauli-Villars regulator is that, as remarked in Ref.~%
\cite{Arriola:reg}, it violates unitarity. This is because for sufficiently
large negative values of $Q^2$ (\mbox{$- Q^2 > 4(M^2 + g_p^2 M^2)$}) the
discontinuity along the cut in meson correlators is negative.

\section{Conclusion}

\label{sec:conclusion}

We have studied the meson propagators in the Nambu--Jona-Lasinio model in
the vacuum and solitonic backgrounds. We have found that the proper-time
regularization introduces extra complex poles to the meson propagators,
which may cause serious complications.
The extra poles, as well as the
non-vanishing contribution of the infinite
semi-circle, forbid the usual deformation of the
energy integration contour which allows one to express the partition
function in terms of the physical modes of excitation of the system. The
unphysical complex singularities, and the infinite semi-circle
piece, have to be accounted for. As a result,
the formal connection to well-known text-book expressions is lost. We
no longer
have the usual dispersion relations for Green's functions, the RPA-equations
for the collective excitations of the system can not be derived in the usual
symplectic form. In general, we cannot use the particle-hole description.

In view of this the naive use of the Cauchy theorem in order to calculate
the RPA-correlation energy for the Nambu--Jona-Lasinio soliton as {\em e.g.}
in Ref. ~\cite{Tu:mass}, is incorrect. The correct way to calculate the
energy and other observables on the one-meson-loop level is to perform
explicitly the integrals over Euclidean four-momenta in the meson
propagators.

Our remarks concerning meson correlators also apply to diquark correlators.
Diquark models have been formulated in
the spirit of the Nambu--Jona-Lasinio model \cite
{Thorsson:di,Vogl:di,Weise:di,Reinhardt:di,Weiss:di,Hellstern:di}. Our
warning is that similar care has to be taken while performing integrals over
diquark momenta, {\em e.g.} when continuing diquark functions from the
Euclidean to the Minkowski space, or solving the Fadeev equations in
quark-diquark systems.

The authors acknowledge the support of the Polish State Committee of
Scientific Research, grant 2~P03B~188~09 (WB), the Maria Sk\l{}odowska-Curie
grant PAA/NSF-95-158, the Bulgarian National Science Foundation,
contract $\Phi$-406 (EN), and the Alexander von Humboldt
Stiftung (GR, WB), which made this collaboration possible.

\newpage
\appendix

\section{Meson propagators in the spectral representation}

\label{app:der}

We expand the first term in Eq.~(\ref{eq:sss}) to second order in mesonic
fluctuations $\delta \phi$ around the reference {\em stationary} state
(vacuum, soliton), which has the meson field configuration $\overline{\phi}$%
. We decompose the $D^\dagger D$ operator as follows: $D^\dagger D = H_0 +
V_1 + V_2$, where
\begin{eqnarray}  \label{eq:shifted}
H_0 & = & - \partial_\tau^2 + {h}^2 \;\;,  \nonumber \\
V_1 & = & - \sum_a \left ( \Gamma_a (\partial_\tau \delta \phi_a) - \{ {h},
\Gamma_a \delta \phi_a \} \right ) \;\;,  \nonumber \\
V_2 & = &\sum_{ab} \Gamma_a \delta \phi_a \Gamma_b \delta \phi_b \;\;.
\end{eqnarray}
Using the expansion
\begin{equation}  \label{eq:expand}
e^{-s(H_0+V)} = e^{-s H_0} - s V e^{-s H_0} + \frac{1}{2} s^2 \int_0^1 d
\beta V e^{-s(1-\beta) H_0} V e^{-s \beta H_0} + ...
\end{equation}
we pick-up the second-order piece of the proper-time effective action
expanded in the mesonic fluctuations $\delta \phi$:
\begin{equation}  \label{eq:secondorder}
S^{F (2)} = \frac{1}{2} Tr \int_{1/\Lambda^2}^\infty ds \left (-V_2 e^{-s
H_0} + \frac{1}{2} s \int_0^1 d \beta V_1 e^{-s(1-\beta) H_0} V_1 e^{-s
\beta H_0} \right)\;\;.
\end{equation}
Now let us use the spectral representation, with states denoted by $\mid
\omega, j \rangle$, etc., where $\omega$ is the frequency variable
(conjugate to $\tau$) and $j$ is as in Eq. \ref{eq:spectrum}. Note that $H_0$
is diagonal in this representation:
\begin{equation}  \label{eq:H0}
H_0 \mid \omega, j \rangle = (\omega^2 + \epsilon_j^2) \mid \omega, j
\rangle \;\;.
\end{equation}
We obtain
\begin{equation}  \label{eq:bubble3}
\langle \vec{x},a \mid K^{-1}(\nu) \mid \vec{y},b \rangle = \delta(\vec{x} -
\vec{y}) \delta_{ab} a^2 + \sum_{jk} \left ( S_{ab}^{jk} + T_{ab}^{jk}
\right ) \;\;,
\end{equation}
where
\begin{eqnarray}  \label{eq:AB}
S_{ab}^{jk} & = & - \frac{1}{2} \left ( q^\dagger_j(\vec{x}) \Gamma_a q_k(%
\vec{x}) \right ) \left ( q^\dagger_k(\vec{y}) \Gamma_b q_j(\vec{y}) \right
) \int \frac{d \omega}{2 \pi} \int_{1/\Lambda^2}^\infty ds e^{-s (\omega^2 +
\epsilon_j^2)} + (a \leftrightarrow b, \vec{x} \leftrightarrow \vec{y}) \;\;,
\nonumber \\
T_{ab}^{jk} & = & \frac{1}{4} \left ( q^\dagger_j(\vec{x}) \Gamma_a q_k(\vec{%
x}) \right ) \left ( q^\dagger_k(\vec{y}) \Gamma_b q_j(\vec{y}) \right )
\int \frac{d \omega}{2 \pi} \left ( (\epsilon_j + \epsilon_k)^2 + \nu^2
\right )  \nonumber \\
& \times & \int_{1/\Lambda^2}^\infty ds\, s \int_0^1 d \beta
e^{-s(1-\beta)((\omega-\nu)^2 + \epsilon_k^2)} e^{-s \beta (\omega^2 +
\epsilon_j^2)} + (a \leftrightarrow b, \vec{x} \leftrightarrow \vec{y}) \;\;.
\end{eqnarray}
Carrying out the $\omega$ integration, changing the integration variable $%
\beta$ to $u = 2 \beta -1$, and noticing that the replacement $(a
\leftrightarrow b, \vec{x} \leftrightarrow \vec{y})$ is equivalent to $(j
\leftrightarrow k,\; u \leftrightarrow -u)$, we get Eq.~(\ref{eq:Ksol}).

\section{Integral of the logarithm of a function.}

\label{app:log}

We recall a useful form of an integral of the logarithm of a function $f$.
First, integrate by parts:
\begin{equation}
I=\frac 12\int_{-\infty }^\infty \frac{d\nu }{2\pi }\log f(\nu )=\left.
\frac 1{4\pi }\nu \log f(\nu )\right| _{-\infty }^\infty -\int_{-\infty
}^\infty \frac{d\nu }{4\pi }\nu \frac{df(\nu )/d\nu }{f(\nu )}\;\;.
\label{eq:logbp}
\end{equation}
Assume that the function $f$ has poles at $\nu =\nu _i^{{\rm pole}}$ and
zeros at $\nu =\nu _j^0$. Then the function $\nu \frac{df(\nu )/d\nu }{f(\nu
)}$ has a set of poles at $\nu =\nu _i^{{\rm pole}}$ with residues $-\nu _i^{%
{\rm pole}}$, and another set of poles at $\nu =\nu _j^0$ with residues $%
+\nu _j^0$. If the surface term vanishes, and if the integral along the
infinite upper (or lower) semi-circle vanishes, then the Cauchy theorem
gives the following result:
\begin{equation}
I=-\frac i2\left( \sum_{j>0}\nu _j^0-\sum_{i>0}\nu _i^{{\rm pole}}\right)
\;\;,  \label{eq:Iom}
\end{equation}
where $j>0$ indicates that ${\cal I}m(\nu _j^0)>0$, and $i>0$ indicates that
${\cal I}m(\nu _i^{{\rm pole}})>0$.

\section{RPA equations.}

\label{app:RPA}

It is convenient to use an {\em orthonormal} basis in the mesonic space,
defined by a set of functions $f_n(\vec{x})$. Each $f_n$ is a four-component
vector in the $\sigma$, $\pi_1$, $\pi_2$, and $\pi_3$ space, {\em i.e.} %
\mbox{$f_n = (f_n^0,f_n^1,f_n^2,f_n^3)$}. We define
\begin{equation}  \label{eq:bas}
K^{-1}_{mn}(\nu) = \sum_{ab} \int d^3x \int d^3y f^a_m(\vec{x}) \langle \vec{%
x},a \mid K^{-1}(\nu) \mid \vec{y},b \rangle {f^b_n}^*(\vec{y}) \;\;.
\end{equation}
Condition (\ref{eq:eigen}) is equivalent to solving the eigenvalue problem
(everywhere below repeated indices are summed over)
\begin{equation}  \label{eq:RPA1}
\lim_{\Lambda \to \infty} K^{-1}_{mn}(\nu)f_n = 0 \;\;.
\end{equation}
Defining the quark-meson overlap coefficients as
\begin{equation}
C^{jk}_m = \sum_a \int d^3x\; q^\dagger_j(\vec x) \Gamma_a q_k(\vec x)
f^a_m(\vec x)
\end{equation}
we can show that the problem~(\ref{eq:RPA1}) is equivalent to the following
set of equations:
\begin{eqnarray}  \label{eq:RPA2}
C^{ph}_n X^{ph} + C^{hp}_n Y^{hp} + a^2 f_n & = & 0 \;\;,  \nonumber \\
(\epsilon_p - \epsilon_h + i \nu) X^{ph} + (C^{ph}_n)^* f_n & = & 0 \;\;,
\nonumber \\
(\epsilon_p - \epsilon_h - i \nu) Y^{hp} + (C^{hp}_n)^* f_n & = & 0 \;\;,
\end{eqnarray}
which can be seen immediately when one eliminates the variables $X_n^{ph}$
and $Y_n^{hp}$ from the above equation --- as a result, the original Eq. (%
\ref{eq:RPA1}) is obtained. One can also eliminate the variables $f_n$ from
Eqs.~(\ref{eq:RPA2}), using the first of Eqs.~(\ref{eq:RPA2}):
\mbox{$ f_n = -
1/a^2 \;(C^{ph}_n X^{ph} + C^{hp}_n Y^{hp}) $}. The second and third
equations become:
\begin{eqnarray}  \label{eq:RPA3}
(\epsilon_p - \epsilon_h + i \nu) X^{ph} - \frac{1}{a^2} \left (
(C^{ph}_n)^* C^{p^{\prime}h^{\prime}}_n X^{p^{\prime}h^{\prime}} +
(C^{ph}_n)^* C^{h^{\prime}p^{\prime}}_n Y^{h^{\prime}p^{\prime}} \right ) &
= & 0 \;\;,  \nonumber \\
(\epsilon_p - \epsilon_h - i \nu) Y^{hp} - \frac{1}{a^2} \left (
(C^{hp}_n)^* C^{p^{\prime}h^{\prime}}_n X^{p^{\prime}h^{\prime}} +
(C^{hp}_n)^* C^{h^{\prime}p^{\prime}}_n Y^{h^{\prime}p^{\prime}} \right ) &
= & 0 \;\;.
\end{eqnarray}
Since the basis $\{f_n\}$ is complete (this is an arbitrary orthonormal
basis, e.g. of plane waves), {\em i.e.}
\mbox{$ \sum_n f_n^a(\vec{x}) f_n^{b*}(\vec{y})
= \delta^{ab} \delta(\vec{x}-\vec{y})$}, we have
\begin{equation}  \label{eq:V}
- \frac{1}{a^2} \sum_n (C^{ij}_n)^* C^{kl}_n = V_{ij,lk} \;\;,
\end{equation}
where $V$ has been defined in Eq.~(\ref{eq:Vijkl}). Using this notation we
may rewrite Eq.~(\ref{eq:RPA3}) as
\begin{eqnarray}  \label{eq:RPA5}
\left (
\begin{array}{cc}
(\epsilon_p - \epsilon_h) \delta_{ph,p^{\prime}h^{\prime}} +
V_{ph,h^{\prime}p^{\prime}} & V_{ph,p^{\prime}h^{\prime}} \\
V_{hp,h^{\prime}p^{\prime}} & (\epsilon_p - \epsilon_h)
\delta_{ph,h^{\prime}p^{\prime}} + V_{hp,p^{\prime}h^{\prime}}
\end{array}
\right ) \left (
\begin{array}{c}
X^{p^{\prime}h^{\prime}} \\
Y^{h^{\prime}p^{\prime}}
\end{array}
\right )  \nonumber \\
= - i \nu \left (
\begin{array}{cc}
\delta_{ph,p^{\prime}h^{\prime}} & 0 \\
0 & - \delta_{ph,h^{\prime}p^{\prime}}
\end{array}
\right ) \left (
\begin{array}{c}
X_n^{p^{\prime}h^{\prime}} \\
Y_n^{h^{\prime}p^{\prime}}
\end{array}
\right )
\end{eqnarray}
This is the symplectic RPA eigenvalue problem, familiar from nuclear
physics, which has the form
\begin{equation}  \label{eq:RPA6}
\left (
\begin{array}{cc}
~A~ & ~B~ \\
~B^*~ & ~A^*~
\end{array}
\right ) \left (
\begin{array}{c}
X \\
Y
\end{array}
\right ) = - i \nu \left (
\begin{array}{cc}
~1~ & ~0~ \\
~0~ & ~-1~
\end{array}
\right ) \left (
\begin{array}{c}
X \\
Y
\end{array}
\right )
\end{equation}
We refer the reader to Refs.~\cite{Rowe,Ring,BRipka} for a detailed
discussion of this symplectic eigenvalue problem. We now want to rewrite the
quantity $\sum_{ph}(\epsilon_p - \epsilon_h)$ differently. From Eq. (\ref
{eq:RPA5}-\ref{eq:RPA6}) we see that
\begin{equation}  \label{eq:tr1}
\sum_{ph}(\epsilon_p - \epsilon_h) = {\rm Tr} A - \sum_{ph} V_{ph,hp} \;\;.
\end{equation}
The trace in this expression may be evaluated using the eigenstates of Eq. (%
\ref{eq:RPA6}), and we get
\begin{equation}  \label{eq:traceA}
\frac{1}{2} Tr A = + \sum_{i>0} \omega_i (\mid X_i \mid^2 + \mid Y_i \mid^2)
+ \frac{\langle P^2 \rangle}{2 M} \;\;.
\end{equation}
We stress that the above result is {\em algebraic} in origin, and holds for
any symplectic problem. The zero-mode piece $\frac{\langle P^2 \rangle}{2 M}$
is generated ``automatically''. In hedgehog models there is also an
angular-momentum zero-mode piece.

Collecting expressions (\ref{eq:loopen},\ref{eq:tr1},\ref{eq:traceA}) and
using the fact that $\mid X_i \mid^2 - \mid Y_i \mid^2 = 1$ \cite
{Rowe,Ring,BRipka}, we get finally
\begin{equation}  \label{eq:loopenfin3}
\lim_{\Lambda \to \infty} E_1 = - \frac{1}{2} \sum_{i>0} \omega_i \mid Y_i
\mid^2 - \frac{\langle P^2 \rangle}{2 M} + \frac{1}{2} \sum_{ph} V_{ph,hp}
\;\;,
\end{equation}

\newpage


\newpage

\renewcommand{\thefigure}{\arabic{figure}}

\begin{figure}[tbp]
\centerline{\psfig
{%
figure=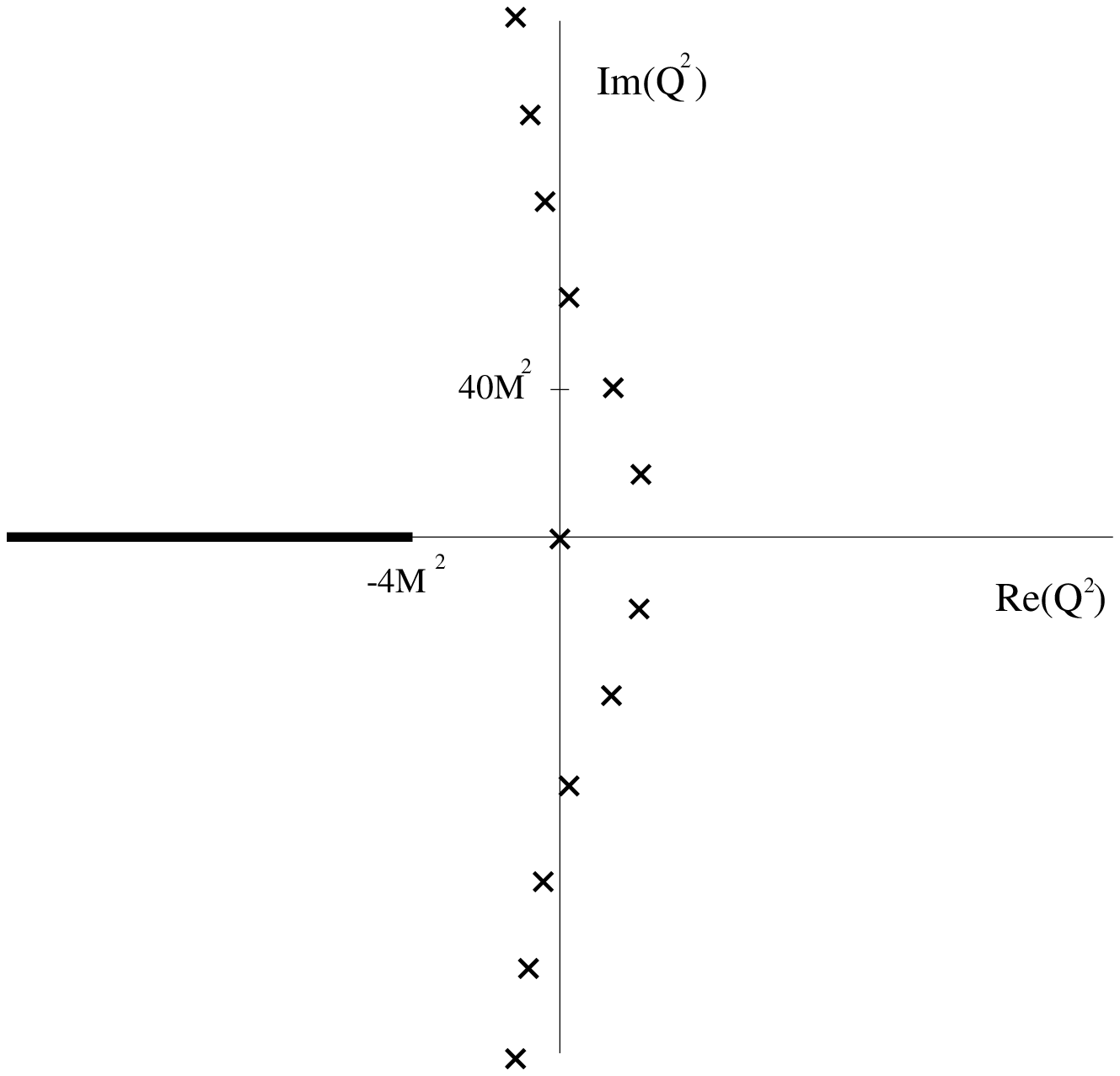,height=11cm,bbllx=100bp,bblly=200bp,bburx=500bp,bbury=600bp,clip=%
}}
\caption{The analytic structure of the pion propagator in the complex
Euclidean momentum plane, $Q^2$. It has the physical pole at $Q^2 = 0$, the $%
q\overline{q}$ production cut from $Q^2 = -4M^2$ to $-\infty$, and
unphysical poles located close to the imaginary axis. The parameters are $%
M=\Lambda$. Note different scales on the real and imaginary axis. The
corresponding plot for the $\sigma$-meson propagator differs only by the
location of the physical pole, which is moved from $Q^2 = 0$ to $Q^2 = -4M^2$%
.}
\label{fig:pionQ2}
\end{figure}

~ \newpage
~

\begin{figure}[tbp]
\centerline{\psfig
{%
figure=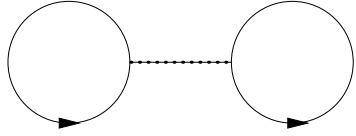,height=5cm,bbllx=10bp,bblly=300bp,bburx=600bp,bbury=500bp,clip=%
}}
\caption{The Hartree diagram. The dotted line is the interaction, which is
local in our case, and carries a factor of $1/N_c$. The diagram is of order $%
N_c$.}
\label{fig:qloop}
\end{figure}

\begin{figure}[tbp]
\centerline{\psfig
{%
figure=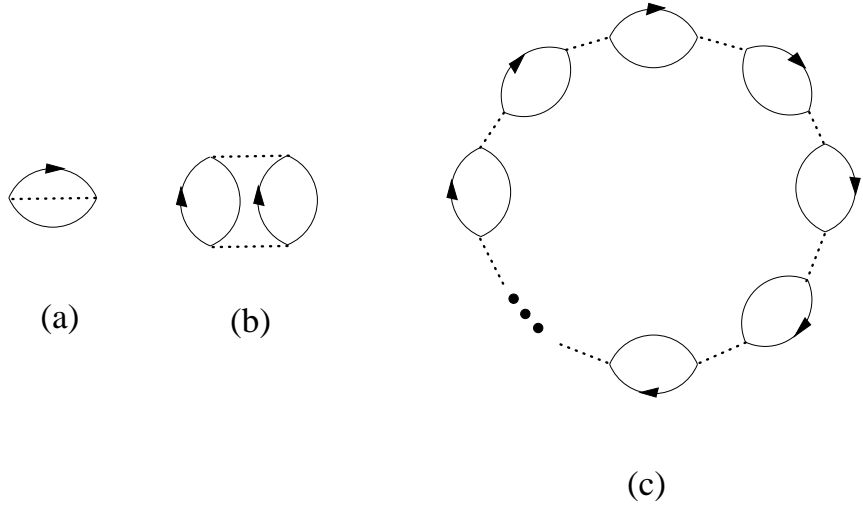,height=8cm,bbllx=30bp,bblly=250bp,bburx=550bp,bbury=550bp,clip=%
}}
\caption{The Fock digram (a), and higher order ring diagrams (b-c). All
these diagrams are of order $N_c^0$.}
\label{fig:mloop}
\end{figure}
\begin{figure}[tbp]
\centerline{\psfig
{%
figure=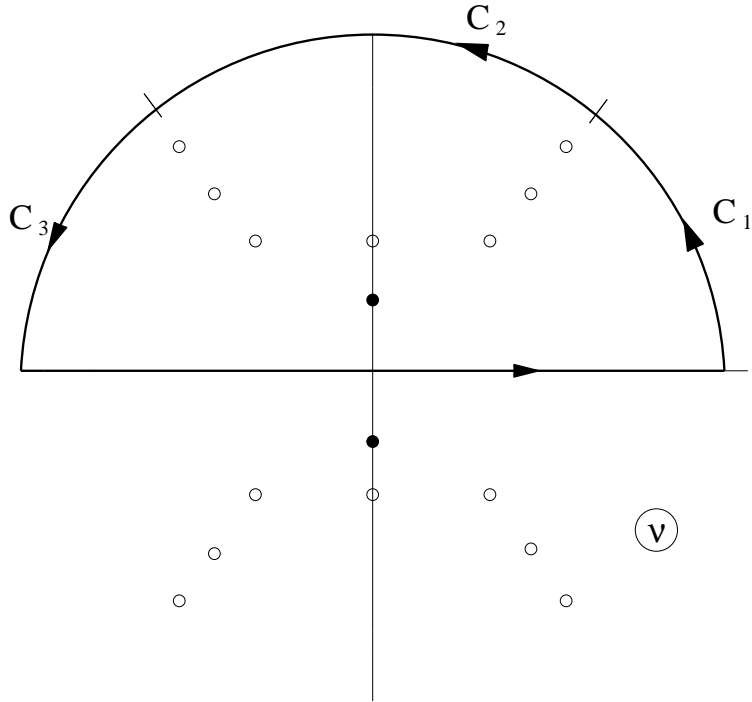,height=11cm,bbllx=50bp,bblly=190bp,bburx=500bp,bbury=600bp,clip=%
}}
\caption{The analytic structure of the function $K^{-1}(\nu)$. It has
poles on the imaginary axis (filled dots), corresponding to particle-hole
excitations, and zeros on the imaginary axis (empty dots), corresponding to
the RPA vibrations. Additionally, it has infinitely many unphysical zeros in
the complex plane off the imaginary axis (empty dots). These zeros become
poles of the meson propagator $K$. The contour integral along $C_3$ diverges
as the radius of the semi-circle goes to infinity.
The sum of pole contributions and the $C_3$-contribution is finite.}
\label{fig:cauchy}
\end{figure}

\end{document}